\documentclass[letterpaper, 10 pt, conference]{ieeeconf}  

\IEEEoverridecommandlockouts                              

\usepackage{graphics} 
\usepackage{epsfig} 
\usepackage{mathptmx} 
\usepackage{times} 
\usepackage{amsmath} 
\usepackage{amssymb}  
\usepackage{subcaption}

\title{\LARGE \bf
Toward Scalable Patient Safety Training: A Prototype for Root Cause Analysis Simulation With AI Virtual Avatars
}

\author{%
  Yuqi Hu$^1$, Qiwen Xiong$^1$, Zhenzhen Qin$^1$, Brandon Watanabe$^2$, Yujing Wang$^1$, Mirjana Prpa$^3$, Ilmi Yoon$^{1*}$\\
  \\
  $^1$Khoury College of Computer Science, Northeastern University, Silicon Valley, San Jose, USA   
  \\$^2$Department of Computer Science, San Francisco State University, San Francisco, USA 
  \\$^3$ Khoury College of Computer Science, Northeastern University, Vancouver, Vancouver, Canada\\
  $^*$Corresponding to: i.yoon@northeastern.edu
}

\begin{document}

\maketitle
\thispagestyle{empty}
\pagestyle{empty}

\begin{abstract}
Patient safety training is essential for preparing healthcare professionals to identify, investigate, and prevent adverse events. However, conventional simulation-based approaches often require substantial faculty time, physical resources, and standardized facilitation. This paper presents a prototype AI-powered simulation platform designed to support more scalable patient safety training through root cause analysis (RCA). The system provides a Unity-based 3D simulation environment, which allows trainees to investigate an ICU adverse event by interviewing five virtual team members represented as AI-powered avatars. Each avatar is driven by a large language model (LLM) agent with role-specific knowledge and variable states of mind. Moreover, emotional text-to-speech and AI-supported facial and body animation enable more realistic and immersive interactions. After completing the simulation, trainees submit a written RCA report and receive rubric-guided formative and summative feedback automatically generated by an LLM-based assessment component. The prototype is built to support patient safety training for healthcare professionals, focusing on skills in communication, investigation, thinking, and analysis, with low recurring instructional burden. We describe the design of the platform, its core technical components, and an RCA case based on a published ICU scenario. This work demonstrates the feasibility of integrating generative AI into immersive simulation for scalable patient safety education.
\end{abstract}


\section{Introduction}
Preventable patient harm remains a persistent challenge in modern healthcare. Since \emph{To Err Is Human} highlighted the scale of medical error and catalyzed the patient safety movement~\cite{iom2000toerr}, national and international organizations have continued to emphasize workforce development as a core strategy for reducing avoidable harm~\cite{who2021gpsap}. Epidemiological studies and subsequent syntheses suggest that adverse events in hospitals are common~\cite{brennan1991harvard}, and modern estimates still indicate substantial harm associated with hospital care~\cite{james2013harms}, including analyses showing that medical error contributes meaningfully to mortality burden~\cite{makary2016medical}. Accordingly, training programs has expanded across professional curricula and continuing education. Global guidance such as the WHO patient safety curriculum supports integrating these competencies into education programs for healthcare professionals~\cite{who2011curriculum}.
A key lesson from patient safety research is that adverse outcomes rarely stem from a single individual mistake; rather, they reflect interacting system factors such as workflow design flaws and communication breakdowns~\cite{reason2000human}. 
Therefore, patient safety training can benefit from structured approaches that focuses on latent conditions rater than surface-level issues~\cite{vincent1998framework,vincent2003understanding}.

Simulation-based education has become a central modality for patient safety training because it enables practice of rare, high-risk events and challenging interpersonal situations without risking patient harm~\cite{gaba2004future,issenberg2005highfidelity,cook2011technology}. Research has shown that simulations can be used to improve clinical performance and selected downstream outcomes~\cite{okuda2009utility,barsuk2009crbsi}, and emphasized its role as an experiential learning strategy~\cite{lateef2010simulation}. Importantly, simulation effectiveness depends not only on scenario fidelity but also on facilitation and debriefing quality~\cite{fanning2007debriefing,rudolph2006debriefing, cheng2014debriefing}.
Despite clear benefits, simulation-based patient safety training faces persistent barriers, including high resource demands (faculty time, standardized patients, simulation centers), variability in facilitation and assessment, and limited scalability for distributed learners. Virtual patients have emerged as a potential solution~\cite{cook2009virtualpatients}, and evidence suggests they can be effective in training programs when appropriately designed and integrated~\cite{kononowicz2019virtualpatients}. Earlier work also explored embodied or dialogue-based simulated standardized patients~\cite{hubal2000vsp} and relational agents that sustain realistic interactions over time~\cite{bickmore2005relational}. However, many prior virtual-agent systems required constrained dialogue or intensive authoring, limiting realism and generalizability~\cite{kenny2007virtualhumans}.

Recent advances in large language models (LLMs) provide a practical foundation for scalable, open-ended conversational agents. Foundation models demonstrated strong general-purpose language capabilities~\cite{brown2020gpt3}, and instruction tuning with human feedback improves controllability and alignment to user intent~\cite{ouyang2022instructgpt}. More capable models have accelerated adoption in education and healthcare contexts~\cite{openai2023gpt4}, including studies examining performance on medical examinations and the potential implications for medical education~\cite{kung2023chatgpt}. Domain-focused medical QA efforts (e.g., Med-PaLM-style systems) further illustrate both progress and remaining limitations in reliable clinical knowledge use~\cite{singhal2024medpalm}. At the same time, evaluations on challenging medical tasks underscore that performance depends heavily on prompting, context, and safeguards~\cite{nori2023gpt4medicine}. These systems also introduce new risks, including hallucinated content in generated text~\cite{ji2023hallucination} and broader ethical and social concerns (e.g., bias, misuse, overreliance)~\cite{weidinger2021ethical}, which need to be carefully considered. Lastly, LLMs has shown promise in automated analysis of student work~\cite{shermis2013aes}, especially when rubric is provided to improve reliability and actionability of feedback~\cite{jonsson2007rubrics}.

In this paper, we introduce an AI-enabled, simulation-based patient safety training system for healthcare professionals, including nurses, clinicians, and administrators, that uses virtual avatars to support interactive practice at scale. The system is built on \textit{Nurse Town}, a Unity-based 3D simulation game for nursing education~\cite{hu2025nursetown}. Using an adverse-event scenario adapted from a published simulation case~\cite{campbell2017simulation}, trainees conduct a structured root cause analysis (RCA) investigation by interviewing ICU team members (LLM-powered virtual avatars), each with role-specific knowledge and perspective, and then complete a written analysis report. RCA is chosen because it provides a good example of a structured patient safety procedure, well suited to simulation-based training. Compared with traditional patient safety training formats, our approach aims to (1) reduce recurring resource needs (e.g., standardized patients, physical simulation centers), (2) emphasize under-addressed investigative communication skills (e.g., eliciting timelines, probing inconsistencies, managing interpersonal dynamics), and (3) provide scalable, rubric-guided feedback on both interviewing process and written analysis quality to reduce ongoing faculty workload.
At a high level, this paper makes the following contributions:
\begin{itemize}
    \item We design and implement an AI-powered 3D simulation platform for patient safety training, featuring virtual avatars with LLM-based dialogue, emotional speech and AI-based facial/body animation.
    \item We design and implement an LLM agent-based assessment component that generates rubric-guided formative and summative feedback on trainee performace.
    \item We present a prototype of the simulation system, using root cause analysis as a concrete case.
\end{itemize}

\section{Related Work}
\subsection{Simulation-Based Patient Safety Training for Healthcare Professionals}
Patient safety education has increasingly emphasized systems thinking, teamwork, and communication skills that enable clinicians to anticipate, detect, and mitigate risk. Global guidance such as the WHO curriculum supports integrating patient safety competencies across health professions programs~\cite{who2011curriculum}. In practice, many institutions also adopt structured curricula (e.g., TeamSTEPPS) to establish shared behavioral expectations and communication tools~\cite{ahrq2014teamstepps}. Evidence in the literature has shown how teamwork failures can contribute to safety risks in fast-paced clinical environments~\cite{manser2009teamwork}. Patient safety training interventions can lead to meaningful improvements in team processes and selected patient outcomes, particularly when facilitated with support mechanisms that promote skill transfer to clinical practice~\cite{weaver2014teamtraining}.

Simulation is, conceptually, a technique for replacing or amplifying real experiences with guided, interactive practice~\cite{gaba2004future}. Simulation has been widely used in patient safety training because it enables deliberate practice with feedback, in a protected environment without risking harm to patients. Existing evidence suggest that simulations, especially high fidelity ones, are associated with effective learning~\cite{issenberg2005highfidelity}. A meta-analysis found that simulation can be an effective educational strategy across different health professions~\cite{cook2011technology}. Simulation has also been positioned as a pathway to improve clinical performance and, in selected domains, downstream patient outcomes~\cite{okuda2009utility}, especially when paired with structured curricula and assessment plans~\cite{lateef2010simulation}. Debriefing, with emphasis on reflective processing, psychological safety, and actionable feedback, is widely viewed as a central component of simulation-based training~\cite{fanning2007debriefing,rudolph2006debriefing}.
Previous studied have found that the structure and quality of debriefing after simulation can meaningfully influence learning outcomes~\cite{cheng2014debriefing}. In addition to center-based simulation, digital simulations and serious games have been explored as scalable alternatives that can increase access to repeated practice and standardization~\cite{graafland2012serious}. More recently, virtual reality and other immersive approaches may have potential to further enhance contextual realism and learner engagement~\cite{pottle2019vr}. Nonetheless, many simulation programs still face barriers including cost, faculty availability, variability in facilitation, and challenges in delivering consistent assessment at scale.

Root cause analysis represents one widely used investigative method in healthcare and is regularly included as part of the broader patient safety training practice. RCA is commonly applied after serious safety events to identify contributing system factors and recommend corrective actions. It has been institutionalized in large healthcare systems such as the Veterans Health Administration~\cite{bagian2001vha} and described in quality improvement practice literature as a structured approach for moving from event description to underlying causes and preventive actions~\cite{rooney2004rca}. In the U.S., oversight bodies (e.g., The Joint Commission) have long tied RCA-style analyses to sentinel event response expectations~\cite{jointcommission2024sentinel}. Despite its importance, research has questioned the effectiveness and efficiency of RCA when poorly executed or inconsistently implemented~\cite{wu2008rcatraining}. To address this gap, educators have explored simulation-based curricula to teach RCA processes and investigative skills more effectively, but these approaches still face resource constraints~\cite{aboumrad2019teachingrca}.

\subsection{LLMs in Healthcare Training and Education}
Simulated patients, embodied by virtual agents, have been studied for decades as a way to provide repeatable, standardized practice for communication, clinical reasoning, and decision-making. Virtual patients are characterized as a flexible instructional modality which is particularly effective when aligned with learning objectives and embedded in broader curricula~\cite{cook2009virtualpatients}. A meta-analysis of evidence indicates that virtual patient interventions can better improve knowledge and skills compared to some traditional approaches, while outcomes depend on design choices such as interactivity, feedback, and integration with instruction~\cite{kononowicz2019virtualpatients}. Earlier systems explored dialogue-driven simulated standardized patients for interview training~\cite{hubal2000vsp} and relational agents designed to support sustained, human-like interaction~\cite{bickmore2005relational}. Broader work on interactive virtual humans for training has emphasized the need to coordinate language, emotion, and non-verbal behavior to achieve believable interactions~\cite{kenny2007virtualhumans}. However, due to limitations of earlier NLP approaches, such systems often only supports scripted interactions, which curbs realism for complex safety investigations and interpersonal dynamics.

The emergence and advancements in LLMs have renewed interest in virtual avatars for patient safety training by enabling open-ended, context-aware dialogues. LLMs have demonstrated strong general language capabilities~\cite{brown2020gpt3}, and more advanced methods such as reinforcement learning from human feedback can further improve instruction-following behavior~\cite{ouyang2022instructgpt}. More capable models have been positioned as general-purpose reasoning and dialogue systems~\cite{openai2023gpt4}, and early studies have explored implications for medical education by evaluating performance on standardized medical exams~\cite{kung2023chatgpt}. Domain-specific evaluations of medical knowledge further highlight both opportunities and limitations of current systems in applications in clinical settings~\cite{singhal2024medpalm,nori2023gpt4medicine}. For patient safety training, LLM-based avatars can be particularly appealing because they can simulate perspectives of different stakeholders (e.g., nurses, physicians, pharmacists, administrators), enabling trainees to practice gathering narratives, reconciling inconsistencies, and communicating professionally under uncertainty.

At the same time, deploying LLM-driven avatars in safety-critical educational contexts raises important concerns. Hallucinated or fabricated statements remain a known failure mode in text generation~\cite{ji2023hallucination}, and ethical and social risks (e.g., bias, privacy concerns, overreliance, and misleading authority cues) warrant careful mitigation strategies~\cite{weidinger2021ethical}. Another challenge is ensuring valid and reliable assessment when learners produce open-ended dialogue and written analyses. Rubric-based assessment is a practical approach for improving reliability, transparency, and feedback usefulness~\cite{jonsson2007rubrics}, and prior research on automated evaluation of written work provides methodological grounding for building structured feedback systems that reduce instructor burden~\cite{shermis2013aes}. Our work builds on this trajectory by integrating LLM-powered avatars into an immersive 3D simulation system for patient safety (RCA) practice and integrating an LLM-based rubric-guided automated feedback system.

\section{Prototype Design and Implementation}
This prototype, built on the foundation of \textit{Nurse Town} \cite{hu2025nursetown}, is a Unity-based 3D simulation game that aims to facilitate RCA training among healthcare professionals. The simulation scenario used in this prototype is adapted from a textbook in nursing education~\cite{campbell2017simulation}. In this prototype, we integrated LLMs, emotional text-to-speech (TTS), and generative animations to create life-like virtual avatars. LLM is also used to assess user performance. GPT-4o \cite{openai2024gpt4o} was selected for its demonstrated capabilities in linguistic and reasoning tasks \cite{b52}.

\begin{figure}[t]
    \centering
    \includegraphics[width=\linewidth]{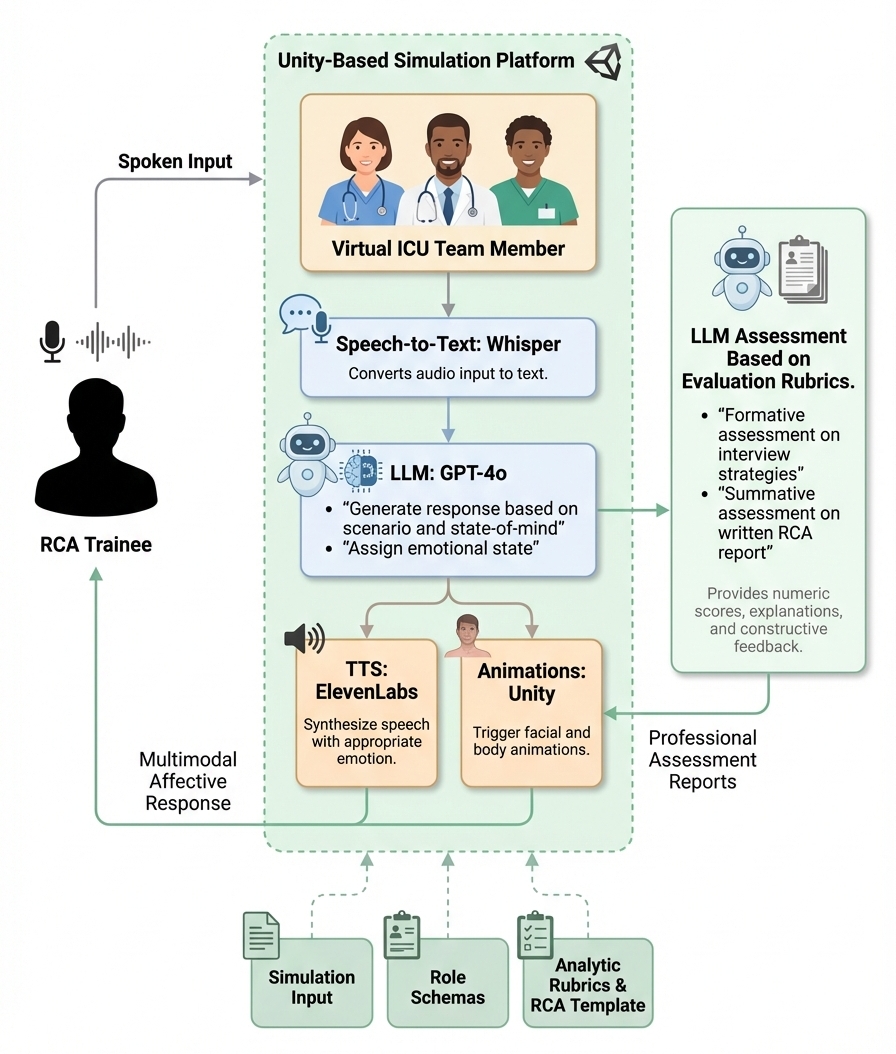}%
    \caption{System architecture of the Unity-based virtual RCA training platform and interactions between components during simulation.}
    \label{fig:system-architecture}
\end{figure}

\begin{figure*}[t]
\centering
\begin{subfigure}{0.9\columnwidth}
    \centering
    \includegraphics[width=\linewidth]{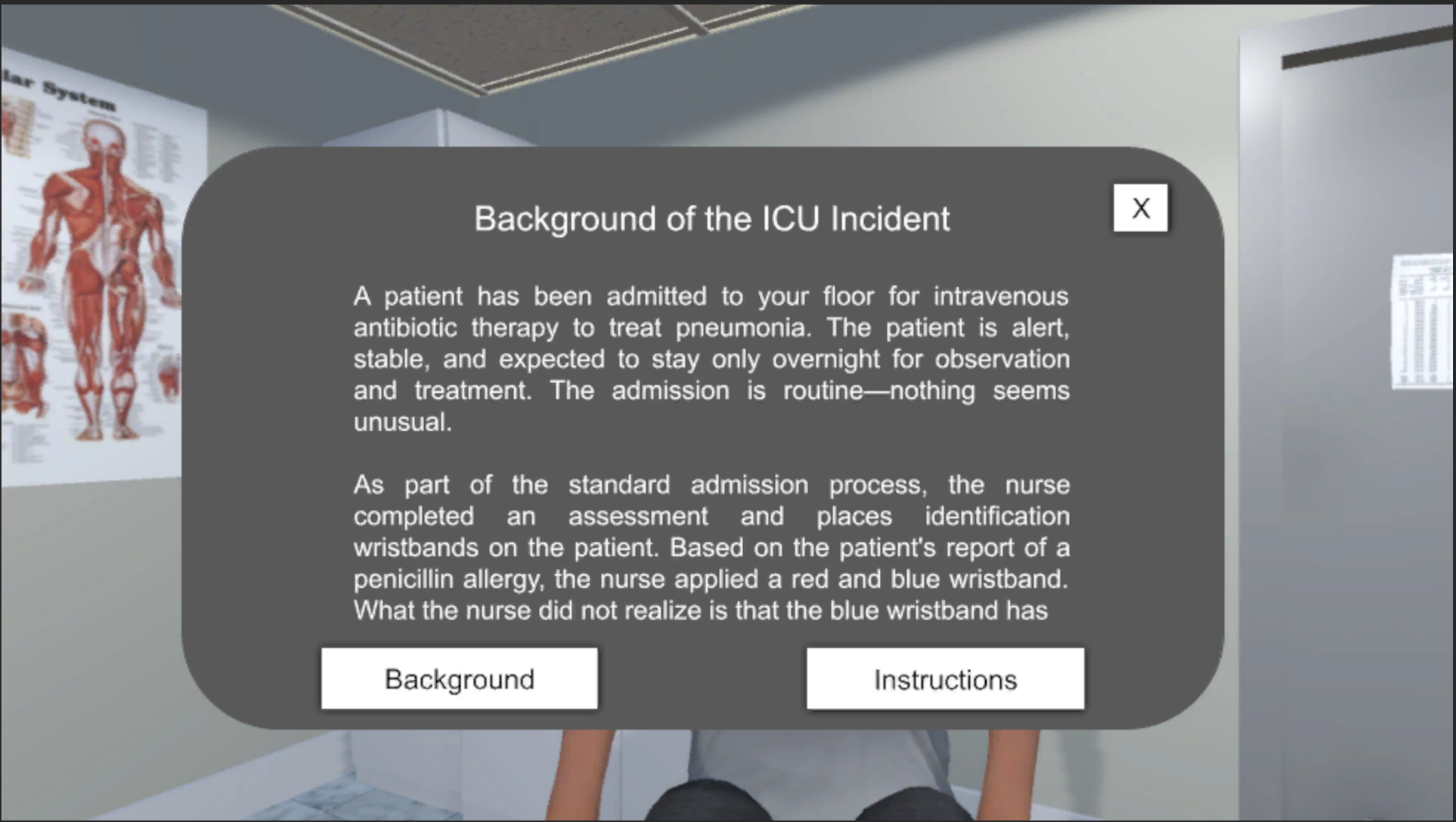}
    \caption{Main interface}
    \label{fig:1a}
\end{subfigure}
\hfill
\begin{subfigure}{0.9\columnwidth}
    \centering
    \includegraphics[width=\linewidth]{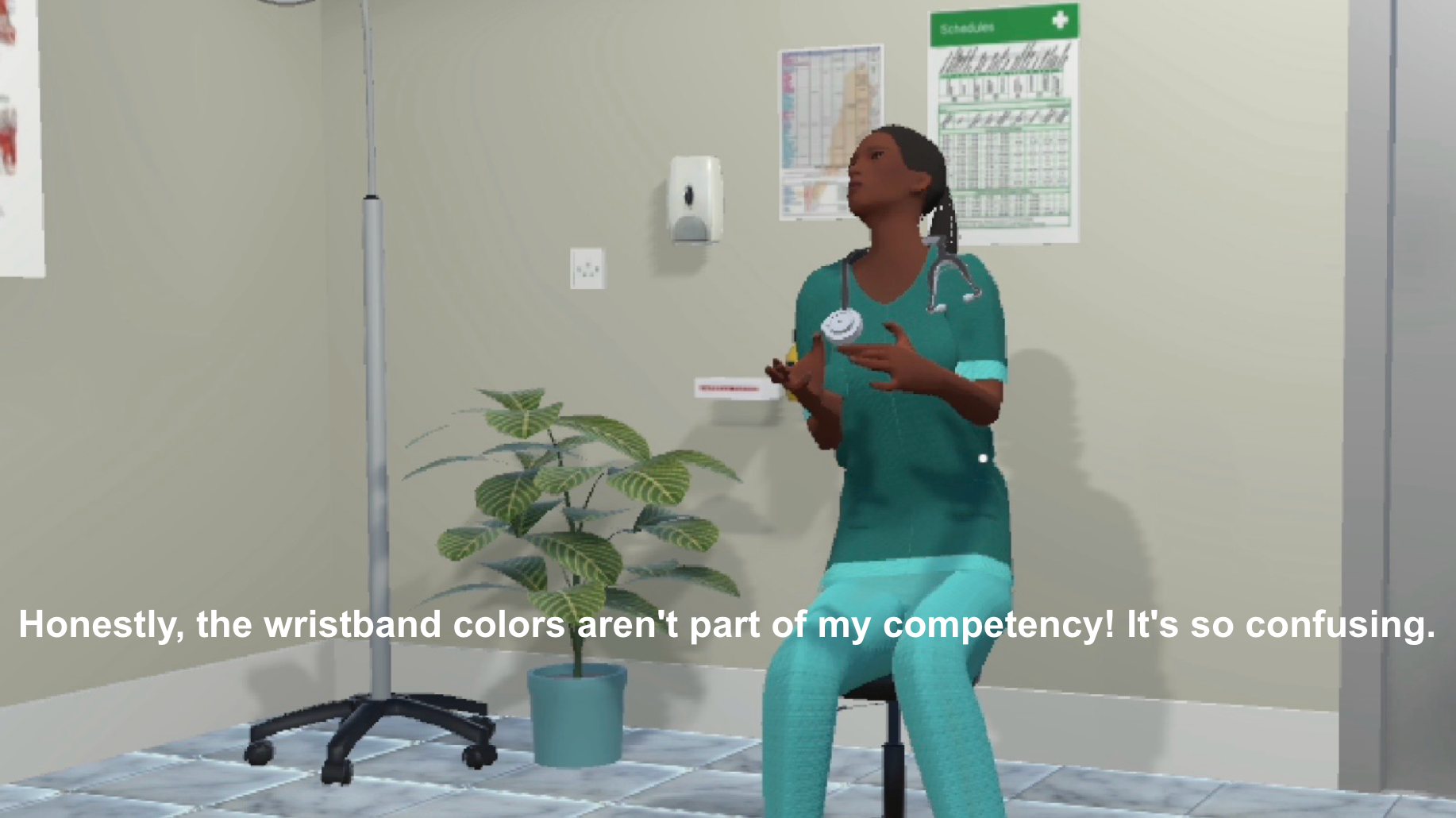}
    \caption{Irritated virtual avatar}
    \label{fig:1b}
\end{subfigure}

\vspace{0.25cm}

\begin{subfigure}{0.9\columnwidth}
    \centering
    \includegraphics[width=\linewidth]{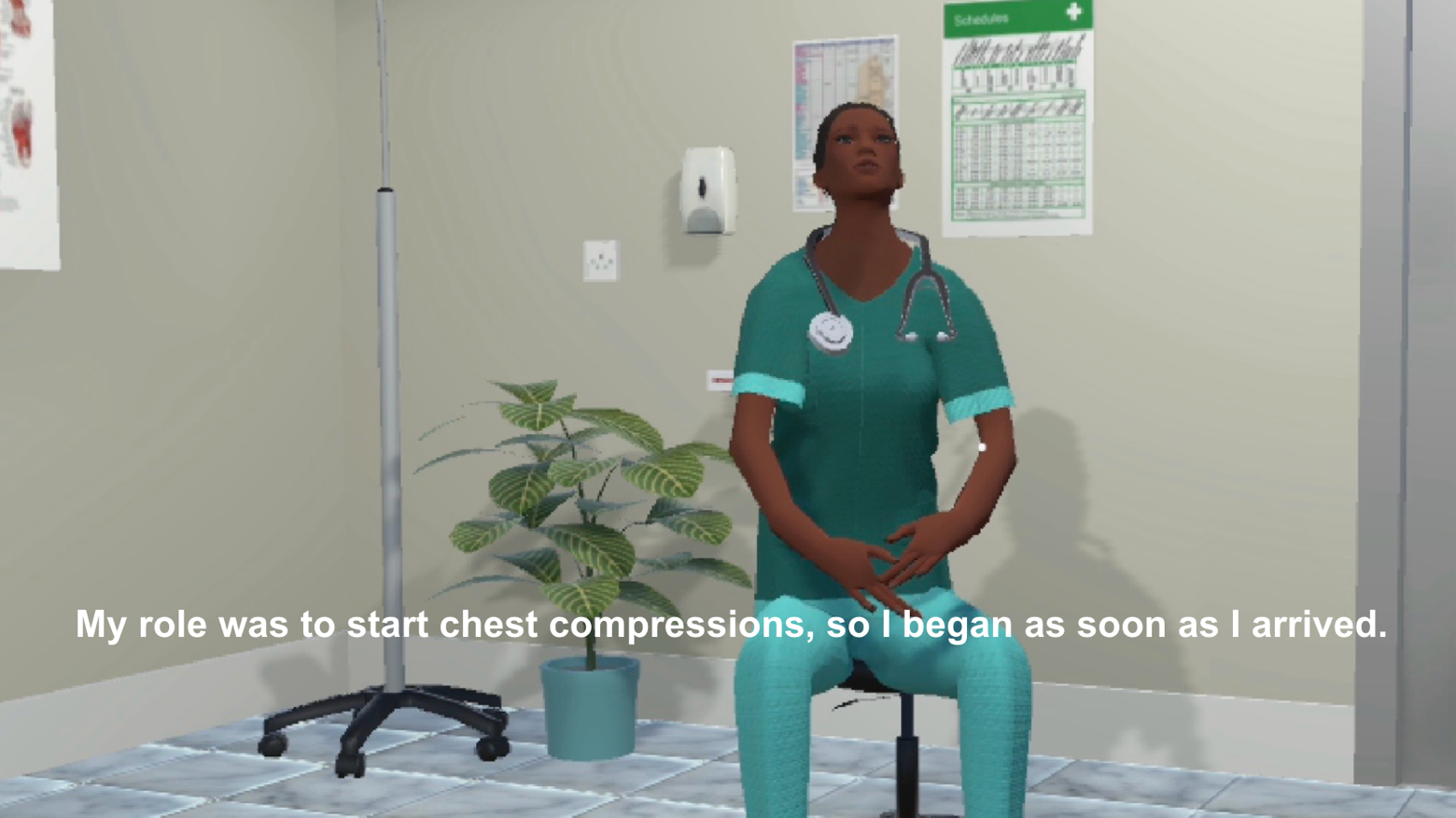}
    \caption{Calm virtual avatar}
    \label{fig:1c}
\end{subfigure}
\hfill
\begin{subfigure}{0.9\columnwidth}
    \centering
    \includegraphics[width=\linewidth]{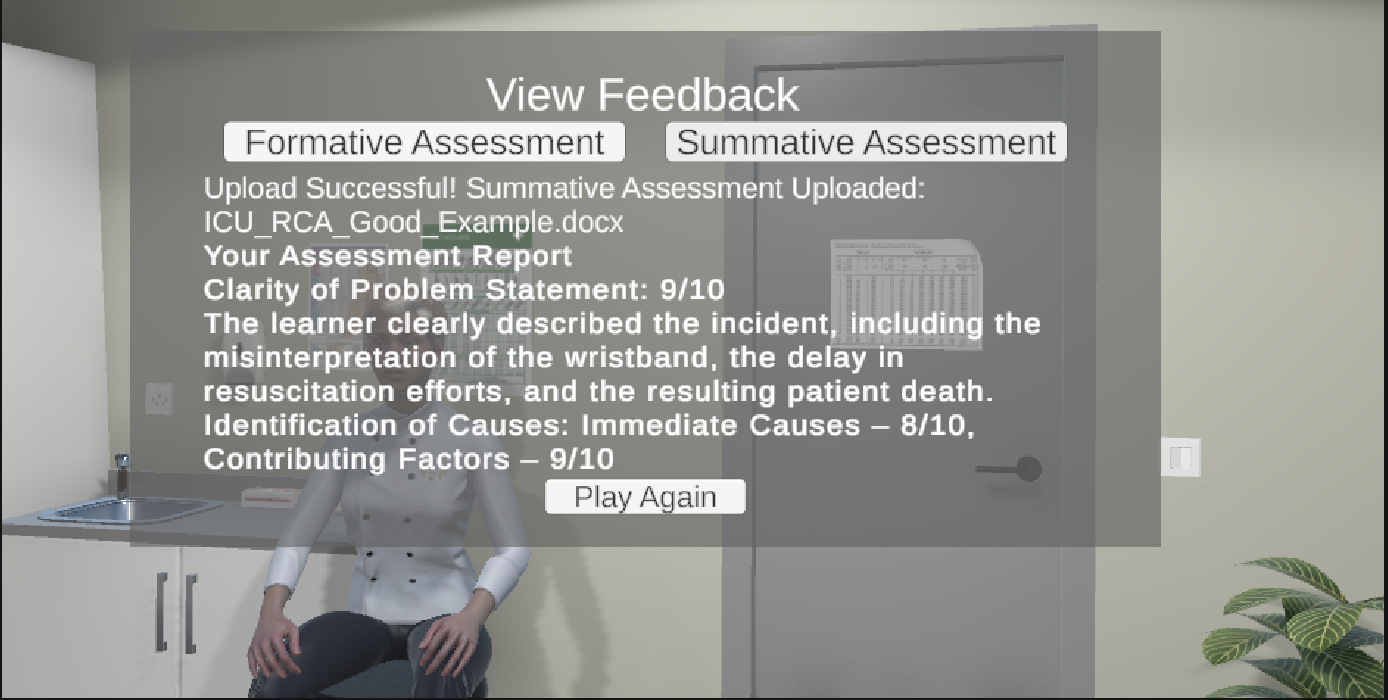}
    \caption{LLM-generated feedback}
    \label{fig:1d}
\end{subfigure}

\caption{Prototype illustrations. Trainees can view background on the case and instructions for conducting the RCA before staring. (top-left). Interviews are conducted via speech-to-speech conversations with virtual avatars, who will have different facial and body animations depending on their emotion state. For example, they will have more dramatic gestures when irritated (top-right) than calm (bottom-left). The session concludes with an LLM-generated assessment report based on the trainee’s RCA submission (bottom-right).}
\label{figure:screenshots}
\end{figure*}

Trainees begin by reviewing the background description of the ICU incident. They are then asked to conduct interviews with the five ICU team members involved in the incident. After all interviews, trainees will complete a written RCA report and receive feedback on the report as well as trainees' interview strategies and techniques. The overall architecture of the Unity-based virtual RCA training platform is shown in Fig.~\ref{fig:system-architecture}, and Figure~\ref{figure:screenshots} demonstrates the essential simulation process with screen captures.

\subsection{Simulation Scenario: ICU Failure Case Study}
The prototype features a textbook scenario for RCA simulation, where a patient died due to communication failures in the ICU team. Specifically, the primary nurse mistakenly placed a blue wristband on the patient because she believed it indicated "allergy," while it actually signals "do not resuscitate" \cite{campbell2017simulation}.
Trainees are tasked to conduct an RCA on this incident. Through interviews with the ICU team members, trainees will investigate, analyze, and report on the case's direct causes (e.g., misinterpretation of wristband color), root causes (e.g., lack of standardized protocols, shift hours that leads to fatigue), and action recommendations (e.g, improved patient admission procedures).
Five virtual characters are available to be interviewed, including the primary nurse, ICU nurse and doctor on the code team, the medical student, and the anesthesiologist. Each character offers a unique perspective based on their roles and experiences in this event. For example, the medical student was not confident in wristband meaning and relied on senior staff for direction.

\subsection{LLM-Based Virtual Avatars}
The prototype includes LLM-powered virtual avatars that represent the five ICU team members involved in the incident. Each avatar is designed to mimic real healthcare professionals, capable of engaging in fluid, context-sensitive conversations. We developed the system instruction prompts through iterative testing and refinement to ensure that the conversations feel authentic and are consistent with the avatars' roles and experiences in the event.

To further enhance realism, each avatar is assigned random states-of-mind (e.g., defensive, confuse, frustrated, etc.) at the start of the interviews. The avatars will respond in different ways accordingly, instead of giving plain factual answers. For example, a "defensive" interviewee may tend to avoid blame and shift responsibility while a "self-reflective" one will openly acknowledge mistakes. Trainees need to adapt their strategies to uncover the full picture. In addition, this diversifies interview experience and mitigates boredom and facilitates continued learning. 

\subsection{Emotional Text-to-Speech}
The system has a text-to-speech component that convert the LLM-generated text responses to life-like speeches.
To facilitate communication, we required the voices to be able to convey nuanced emotional states such as fatigue, hesitation, and urgency. 
Therefore, we adopted ElevenLabs’ voice design and cloning tools ~\cite{elevenlabs2025}, which allowed us to define vocal traits tailored for each character. For example, the code team doctor would have a calm and commanding voide with steady pacing. In addition, ElevenLabs can automatically recognize emotional cues from text input and render speeches accordingly. 
Its integration enabled dynamic emotional responses that align with character roles and state-of-minds.

\subsection{AI-Powered Character Animations}

\begin{figure*}[t]
  \centering
  \includegraphics[width=0.65\linewidth]{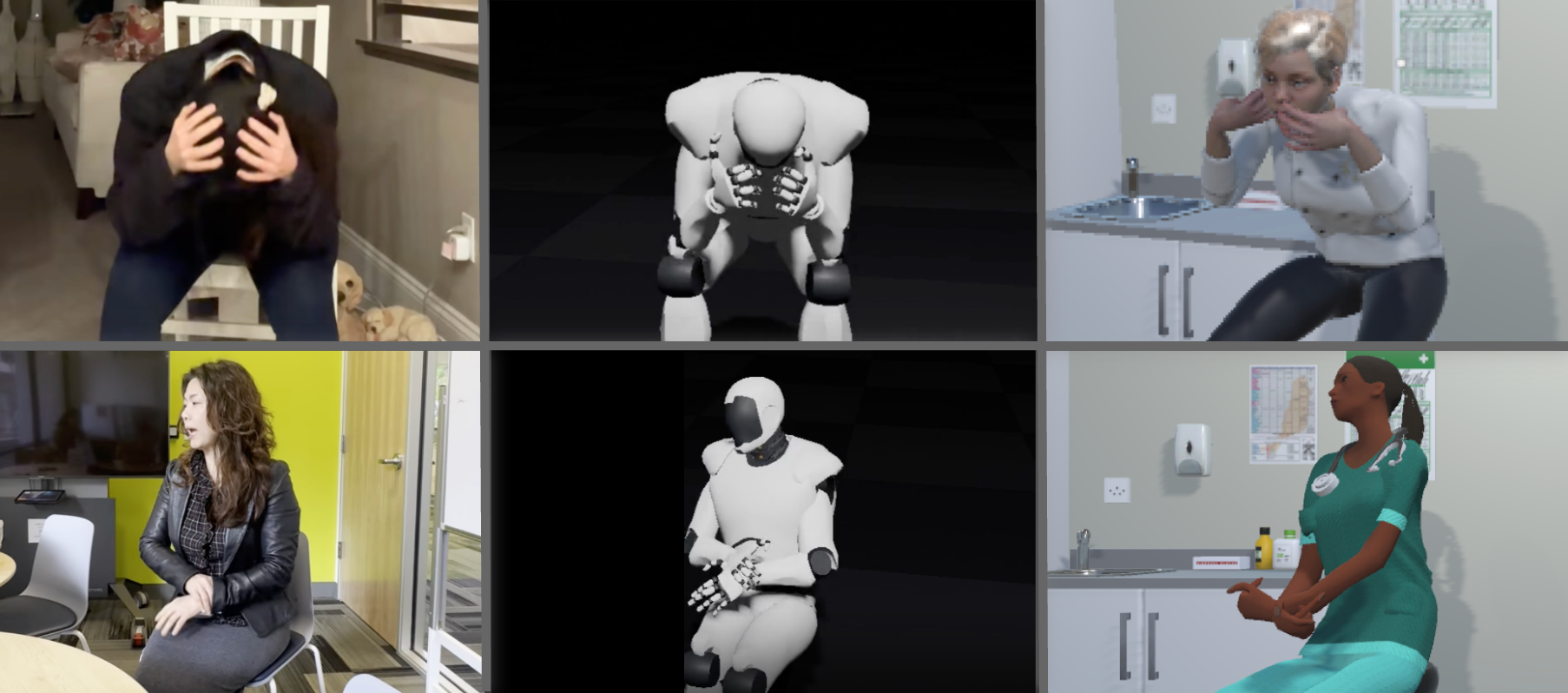}
  \caption{The process of generating customized animations. Left: motions performed by real humans. Middle: skeletal animations generated by Move.ai. Right: animations applied to 3D avatars in Unity.}
\label{figure:motion}
\end{figure*}

AI tools were utilized to enable character animations that are natural, realtime and context-appropriate.
First, to complement ready-made animations acquired from Adobe's Mixamo ~\cite{mixamo}, we created a customized animation library with Move.Ai ~\cite{moveai}. We recorded custom animations through motion capture performed by real humans. These recordings were processed using Move.ai ~\cite{moveai} to convert live-action performances into high-quality 3D animation data. The resulting animations were then refined in Blender ~\cite{blender} to ensure smooth transitions and realistic movements before being integrated into the simulation game. Figure ~\ref{figure:motion} illustrates this process.

During the interviews, an LLM module analyzes the avatar's text responses for emotional cues (e.g., sadness, frustration) and triggers the corresponding animation from the library to align with the emotional state. In addition, we included a set of idle animations, such as gentle swaying, head tilts, or shoulder shrugs, to be played randomly throughout the interactions. These subtle movements further improve naturalness and distinguish the virtual avatar from common game characters.

\subsection{LLM-Based Assessment of Student Performance}
The LLM-based assessment system offers a dual-layer evaluation process. In the formative assessment, the trainee's performance is continuously monitored through their interactions with the virtual avatars. As they conduct interviews, the system analyzes the quality of their questions, their ability to follow up on vague or inconsistent responses, and their skill in identifying key issues during the conversation. When all interviews are completed, the LLM will provide formative feedback on their interview strategies and communications skills to help trainees refine their approaches.

For the summative assessment, trainees will be provided with an RCA template developed according to Center of Medicare and Medicaid Services' official guidance \cite{cms2013rca}, including identification of contributing factors, root causes, and corrective actions. The submission is then examined by LLM on articulation of the problem, identification of direct and root cause, and action recommendation. Overall structure and clarity of the report are also considered. These assessment provide feedback that help trainees develop and strengthen their interviewing techniques and their skills to analyze and synthesize complex information to form a structured, well-argued RCA report.

\section{Discussion}
This work demonstrates the feasibility of using an AI-enabled simulation platform to support patient safety training through an RCA instance. The prototype offers an immersive, interactive environment for practicing interviews, information gathering, and structured analysis. Informal review of avatar dialogues and assessment outputs by individuals with healthcare and educational backgrounds suggested that the interactions were plausible and the feedback was useful for training. Compared with traditional simulation, this approach may reduce recurring resource demands while still supporting repeated practice and individualized feedback.

A key strength of the system is its emphasis on realism and immersion in patient safety training. By combining LLM-powered dialogue, emotional text-to-speech, and AI-supported facial and body animation, the platform creates a multimodal simulation experience that more closely resembles real interpersonal encounters in clinical settings. Rather than interacting with static cases or text-only interfaces, trainees engage with virtual avatars that speak, respond, and express affect in ways intended to increase contextual realism.
The automated assessment component is another important contribution. In many simulation programs, high-quality feedback depends on substantial faculty time and debriefing expertise, which can limit scalability. In this prototype, LLM-based assessment provides formative feedback on interview performance and summative feedback on written RCA reports using explicit rubric criteria.

More broadly, the work points to the potential of AI-powered simulations for scalable patient safety education. Although simulation is widely valued in healthcare training, access is often constrained by cost, faculty availability, and facility requirements. A virtual platform may provide a more accessible and standardized option, especially for distributed learners or programs seeking more frequent practice opportunities.

\subsection{Limitations}
Several limitations should be noted. First, the system cannot fully capture the complexity of real RCA processes, which are shaped by organizational culture, hierarchy, interpersonal tension, and subtle nonverbal cues. The platform should therefore be viewed as a supplement to, rather than a replacement for, in-person training.
Second, avatar affect is still relatively static. Although each avatar begins with a particular state of mind, these emotional states do not yet change dynamically during the interview. In real investigations, an interviewee’s tone and openness often shift over time.
Third, the current prototype includes only one case scenario. Broader educational value will require a larger set of cases spanning different event types, clinical settings, and professional roles.
Finally, this work does not yet provide evidence of learning gains, transfer to practice, or patient safety impact. These outcomes require future empirical study.

\subsection{Future Work}
Future work should begin with formal evaluation of usability, acceptability, perceived realism, and educational effectiveness. Studies should examine whether the system improves RCA knowledge, interviewing strategies, report quality, and learner confidence, and whether these gains transfer to practice. Another priority is expanding the scenario library. A more diverse set of cases would improve generalizability, support progressive learning, and better reflect the range of patient safety events encountered in healthcare.

Future development should also make avatars more adaptive. Dynamic emotional responses and more nuanced interpersonal behavior could improve realism and strengthen communication training. Multi-user functionality may also support collaborative RCA practice across professions. The assessment component also needs further validation. Future studies should compare LLM-generated feedback with expert ratings and examine how AI can best support, rather than replace, instructor judgment.
Finally, implementation research is needed to understand how this type of system can be integrated into existing healthcare training programs. Questions of adoption, oversight, and cost-effectiveness will be important for real-world use.





\section*{ACKNOWLEDGMENT}
We appreciate help from Kunyi Shi and Northeastern University, Silicon Valley.


\end{document}